\documentstyle[12pt]{article}
\begin{document}
 
\baselineskip=20pt
  
\begin{titlepage}

\begin{flushright}
LBL-38475
\end{flushright}
\vspace{0.6in}

\begin{center}
{\Large\bf Effective mass of phi mesons at finite temperature}

\vspace{0.3in}
Chungsik Song\\
{\it Nuclear Science Division, MS70A-3307}\\
{\it Lawrence Berkeley Laboratory, Berkeley, CA 94720, USA}\\
\vspace{0.3in}

{\bf Abstract}
\end{center}
The effective mass of phi meson at non-zero temperature is re-examined with an
effective chiral Lagrangian. 
We find that the phi mass decreases with temperature but the effect is
small compared to the result obtained from calculations using QCD sum rules.
The leading contributions come from kaon loop corrections but vector meson
contributions are also important as temperature increases. 
We discuss consequences of these changes to the phenomena of chiral phase
transition in hot hadronic matter. 

\end{titlepage}


It is expected that at very high temperatures and/or densities
hadronic matter undergoes a phase transition/crossover into a plasma phase 
composed of weakly interacting quarks and gluons \cite{qgp}.
It may be possible 
to produce and observe this new phase of hadronic matter
in relativistic collisions between two heavy nuclei.
Experiments are now being carried out at BNL AGS and at CERN SPS, and  
in the future there will be experiments for which one anticipates the
production of hadronic matter with very high energy-densities enough to form 
a quark-gluon plasma. In the plasma phase of hadronic matter
the spontaneously broken chiral symmetry would be restored.
The restoration of the chiral symmetry can be characterized by 
the vanishing of the quark condensate, which is known as the order parameter 
of the phase transition. 
It has been shown by the numerical simulation of QCD 
formulated on the lattice \cite{lattice} 
and also from some model calculations \cite{gl} that the quark condensate 
diminishes with rising temperature. 

Effective masses of vector mesons at non-zero temperature 
have been studied in various ways 
since it was suggested that they are closely 
related to the phenomena of 
chiral symmetry restoration in hot hadronic matter
\cite{pisaski}.
Moreover, these changes in the vector meson masses can be 
observed in the dilepton spectrum
from hot dense matter produced in high energy nucleus-nucleus collisions.
In lattice simulations, screening masses of these vector mesons,
related to the imaginary-time response function, are measured to study the
structure of the hot matter. It has been suggested that the restoration of the
chiral symmetry are reflected by the degeneracy of the screening masses of
the expected chiral multiplets \cite{screen}. However, there is no general
proof that the screening masses are directly related to 
the physical masses of hadrons.
Effective masses of hadrons can also be defined as the pole
positions of the real-time propagator in the medium.
This pole mass, which is relevant to the peak position in dilepton spectrum,
has been calculated by using QCD sum rules \cite{sum,dey,lee} 
and in effective Lagrangian approaches \cite{gale,song1,haglin,song2}.  

Gale and Kapusta \cite{gale} have calculated vector meson masses at finite 
temperature ($T$) using an effective model with (charged) 
pseudoscalar mesons and (neutral) vector mesons.  
They show that effective masses of vector mesons increase with
temperature at leading order of $T^2$.
Extensive study for phi mesons at finite temperature has been done 
in the $SU(3)$ limit and shows the same result \cite{haglin}.
Even though this model is very simple and satisfies the 
vector meson dominance assumption, however, the model does not satisfy the 
chiral symmetry which is an important feature in hadron properties. 
Actually, we have showed that the $\rho$ mesons mass is not changed in lowest
order of temperature, $T^2$, with a chirally invariant effective model. 
The $\pi-\pi$ tadpole diagram contribution which is
responsible for the increase of the effective mass is exactly 
canceled in this model \cite{song2}. 
The same result also has been obtained from the model independent 
calculation based on chiral symmetry and current algebra \cite{dey}. 
This implies that chiral symmetry is important for the properties of
hadrons at finite temperature and/or density.  

The purpose of this paper is to calculate the effective mass of phi meson 
at finite temperature with an effective Lagrangian which preserves 
chiral symmetry. 
Properties of phi mesons in hot and dense matter have been interested
since they affect the phi meson production in high energy nucleus-nucleus
collisions, which probes the enhancement of
strangeness in quark-gluon plasma phase \cite{phi}. 
To estimate the production rate we
conventionally associate the lepton yields with the total particle abundance,
assuming an invariant branching ratio of the leptonic decay to the hadronic
channel. 
However, a small change of phi mass in hot hadronic matter can easily 
affect the decay modes and the result for dilepton yields.
This is so because the phi mass is very close to 
threshold for the decay into two kaons, the dominant hadronic decay processes 
of phi mesons.
A change of phi mass in hot hadronic
matter also has been suggested as a possible probe for the phase transition
in hot matter \cite{ko}. 
In case of a first order phase transition with a long-lived mixed phase, we
can have double phi peaks in dilepton spectra. 
The second phi peak is from the decay of phi mesons in the mixed phase, which
have reduced masses as a result of partial restoration of chiral symmetry.


Hidden local symmetry (HLS) is a natural framework for 
describing the vector mesons
in a manner consistent with chiral symmetry of QCD \cite{hls}. 
For $SU(N)\times SU(N)$ symmetry, it is constructed with two 
$SU(N)$-matrix valued variables $\xi_L(x)$ and $\xi_R(x)$, which transform 
as $\xi_{L,R}(x) \rightarrow \xi'_{L,R}(x)=h(x) \xi_{L,R}~ g^\dagger_{L,R}$ 
under $h(x)\in [SU(N)_V]_{\rm local}$ and 
$g_{L,R}\in[ SU(N)_{L,R}]_{\rm global}$.
Vector meson fields $V_\mu$  are introduced as the gauge field of the local 
symmetry while the photon fields can be regarded as an external gauge field of
the global symmetry. The HLS Lagrangian yields
at tree level a successful phenomenology for pions and $\rho$ mesons.

We consider the $G_{global}\times H_{local}$ "linear model" with
$G=SU(3)_L\times SU(3)_R$ and $H=SU(3)_V$. 
The chirally invariant Lagrangian is given by 
\begin{eqnarray} 
{\cal L} &=& {\cal L}+\Delta{\cal L}
 \nonumber \\ [12pt] 
         &=& -{1\over 4}f^2 {\rm tr} \left[
             ( {\cal D}_\mu \xi_L \cdot \xi_L^\dagger 
             - {\cal D}_\mu \xi_R \cdot \xi_R^\dagger)^2 
               (1 + \xi_L \epsilon_A \cdot \xi_R^\dagger
                  + \xi_R \epsilon_A \cdot \xi_L^\dagger)\right]
 \nonumber \\ [6pt] 
         & & -\quad {1\over 4}af^2 {\rm tr} \left[
             ( {\cal D}_\mu \xi_L \cdot \xi_L^\dagger 
             + {\cal D}_\mu \xi_R \cdot \xi_R^\dagger)^2  
               (1 + \xi_L \epsilon_V \cdot \xi_R^\dagger
                  + \xi_R \epsilon_V \cdot \xi_L^\dagger)\right]
 \nonumber \\ [6pt] 
         & & +\quad {\cal L}_{\rm kin} ( V_\mu)\,,
\end{eqnarray}
where $a$ is an arbitrary constant and the value $a=2$ is used in the paper.
The covariant derivative ${\cal D}^\mu \xi_{L(R)}$ is given by 
\begin{eqnarray} 
{\cal D}^\mu \xi_{L(R)}= \partial^\mu \xi_{L(R)}+ ig \xi_{L(R)} V^\mu\,,
\end{eqnarray}
and $f$ can be identified with the pion or kaon decay constant. 
Here we add an explicit $SU(3)_V$ symmetry breaking terms, $\Delta{\cal L}$, to
give different masses for $\rho$ and $\phi$ mesons. 
The matrix $\epsilon_{V(A)}=(0,0,c_{V(A)})$ 
where $c_V$ and $c_A$ are constants. These
constants are inferred from the masses of vector mesons and
the ratio of decay constant of $\pi$ to koans at zero temperature \cite{su3}.

In unitary gauge, $\xi^\dagger_L=\xi_R=\xi$, the Goldstone boson fields can
be written as a $3\times 3$ unitary matrix 
\begin{equation} \xi = \exp {i\Phi\over f}, 
\label{eq:1}
\end{equation}
where
\begin{equation}\label{eq:2}
\Phi = {1\over\sqrt{2}}\left[
\begin{array}{ccc}
{\pi^0\over\sqrt{2}} + {\eta_8\over \sqrt{6}} & \pi^+ & K^+ 
\\ \pi^- & -{\pi^0\over \sqrt{2}} + {\eta_8\over \sqrt{6}} & K^0 \\ 
K^- & \overline K^0 & - {2\eta_8\over \sqrt{6}} 
\end{array} \right].
\end{equation}
The $\eta_8$ is not the physical $\eta$ meson 
because of the mixing with a
$SU(3)$ singlet pseudoscalar $\eta_1$. However, this difference will be
neglected since the mixing effect is not important in our conclusion.
Throughout this paper $\eta_8$ will be regarded as physical one 
with mass which is given by Gell-Mann-Okubo relation 
\begin{equation}
m^2_{\eta_8}={1\over3}(4m_K^2-m_\pi^2).
\end{equation}   
The vector meson fields can be written as a 3$\times$3 nonet matrix with 
an ideal mixing between $\phi$ and $\omega$ mesons
\begin{equation}\label{eq:3}
V_\mu = {1\over\sqrt{2}}\left[
\begin{array}{ccc}
{\rho_\mu^0 \over \sqrt{2}} + {\omega_\mu \over \sqrt{2}} & \rho_\mu^+
&K_\mu^{*+}\\ \rho_\mu^- & -{\rho_\mu^0 \over \sqrt{2}} + {\omega_\mu \over
\sqrt{2}} & K_\mu^{*0}\\ K_\mu^{*-} & \overline K_\mu^{*0} & \phi_\mu
\end{array}
\right]\, .
\end{equation}


We define the effective mass of phi meson in medium as a pole
position of the propagator.  The in-medium propagator ${\cal D}^{\mu\nu}$
can be written with free propagator ${\cal D}_0^{\mu\nu}$ 
and self-energy $\Pi^{\mu\nu}$ as   
\begin{equation}
{\cal D}_{\mu\nu}^{-1}={\cal D}_{0\mu\nu}^{-1}+\Pi_{\mu\nu}.
\end{equation}
Thus the self-energy includes all medium effects on the propagation of phi
mesons in hot hadronic matter and changes in their effective mass.
The general form of the self-energy is given by \cite{kapusta}
\begin{equation} 
\Pi_{\mu\nu}=\alpha A_{\mu\nu}+\beta B_{\mu\nu}
            +\gamma C_{\mu\nu}+\delta D_{\mu\nu},
\end{equation}
where $A, B, C$, and $D$ are independent covariant tensors
\begin{eqnarray}
A_{\mu\nu} &=& g_{\mu\nu}-{1\over \vec k^2}
               \left[ k_0( n_\mu k_\nu+n_\nu k_\mu)
                     -k_\mu k_\nu -k^2n_\mu n_\nu\right],\cr
B_{\mu\nu} &=& -{k^2\over \vec k^2}\left(n_\mu-{k_0 k_\mu \over k^2}\right)
                                   \left(n_\nu-{k_0 k_\nu \over k^2}\right),\cr
C_{\mu\nu} &=& -{1\over\sqrt{2}\vert\vec k\vert}
                \left[ \left(n_\mu-{k_0 k_\mu \over k^2}\right)k_\nu
                      +\left(n_\nu-{k_0 k_\nu \over k^2}\right)k_\mu\right],\cr
D_{\mu\nu} &=& {k_\mu k_\nu \over k^2}.
\end{eqnarray}
Here $n_\mu$ specifies the rest frame of the hot matter.   
$\alpha, \beta, \gamma$, and $\delta$ are given by 
\begin{eqnarray}
\delta &=& {1\over k^2} k^\mu k^\nu \Pi_{\mu\nu},\cr
\gamma &=& {\sqrt{2}\over\vert\vec k\vert}(k^\mu\Pi_{\mu 0}-k_0\delta),\cr
\beta &=& -{1\over\vec k^2}\left( k^2\Pi_{00}
                            -\sqrt{2}\vert\vec k\vert k_0\gamma
                            -k_0^2\delta\right),\cr
\alpha &=& {1\over2}(\Pi^\mu_\mu-\beta-\delta).
\end{eqnarray}
The in-medium propagator also can be written in the same form 
\begin{equation}
{\cal D}_{\mu\nu}=  {1\over a} A_{\mu\nu}
                  + {2\over c^2+2bd}(d B_{\mu\nu}-c C_{\mu\nu}+b D_{\mu\nu}),
\end{equation}
with 
\begin{eqnarray}
a &=& k^2-m^2+\alpha,\cr
b &=& k^2-m^2+\beta,\cr
c &=& \gamma,\cr
d &=& -m^2+\delta.
\end{eqnarray}

The self-energy of the phi meson in hot hadronic matter is 
obtained from the diagrams shown in Fig.~1. 
With an exact $SU(3)_V$ symmetry the dominant contribution comes from  
the first diagram Fig.~1.a and is given by 
\begin{equation}
\Pi^{(a)}_{\mu\nu}(k) = -2g^2_{\phi KK} 
                       T\sum\int{d^3l\over(2\pi)^3}
     {(2l_\mu+k_\mu)(2l_\nu+k_\nu)\over(l^2-m_K^2)((l+k)^2-m_K^2)}, 
\end{equation}
where 
$g_{\phi KK}=g(1+2c_V)/(\sqrt{2}(1+c_A))$.
In the chiral limit, $m_K\to 0$, 
it is  proportional $(T^2/f_K^2)^2$.
This is consistent with the 
result based on the current algebra, which show that vector meson
mass is not changed in the leading order of temperature, ${\cal O}(T^2/f_k^2)$.
However, we cannot neglect the mass of kaon, $m_K\sim 500$ MeV, even though it
might be a good approximation for pions.
With finite kaon mass the medium effect cannot be expanded  
by powers of temperature. Instead we use the fact that kaon density is
very small at temperatures we are interested in. In this case we can expand 
by the number of loops and include one loop diagrams as leading terms 
for the calculation of the phi effective mass.   

When we include an explicit $SU(3)_V$ symmetry breaking term, 
$\Delta {\cal L}$,
we have vector-vector-pseudoscalar-pseudoscalar couplings
which are proportional to $c_V$
\begin{eqnarray}
{\cal L}_{\phi\phi KK} &=& - {1\over2}g^2{c_V\over 1+c_A} 
                             (K^0\bar K^0+K^+K^-)\phi_\mu\phi^\mu,
\nonumber\\[6pt]
{\cal L}_{\phi\phi\eta\eta} &=& - {2\over3}g^2{c_V\over 1+(4/3)c_A} 
                                  \eta\eta\phi_\mu\phi^\mu.
\end{eqnarray}
From these couplings we have the contributions (Fig.~1.b)
with small coefficients proportional to
symmetry breaking effect
\begin{eqnarray}
\Pi^{(b)}_{\mu\nu}(k)&=&  2 g^2{c_V\over 1+c_A} g_{\mu\nu}
                         T\sum\int{d^3l\over(2\pi)^3}{1\over(l^2-m_K^2)}, 
\nonumber\\[6pt]
                     && + {4\over3}g^2{c_V\over 1+(4/3)c_A} g_{\mu\nu}
                         T\sum\int{d^3l\over(2\pi)^3}{1\over(l^2-m_{\eta}^2)}. 
\end{eqnarray}

We also include the Wess-Zumino term which is an anomalous part of the
Lagrangian. It contains the vector-vector-pseudoscalar meson interactions
\begin{equation}
{\cal L}_{WZ}=2g_{WZ}\epsilon^{\mu\nu\lambda\sigma}
{\rm tr}\biggl[\partial_\mu V_\nu\partial_\lambda V_\sigma\Phi\biggr],  
\end{equation}
where
\begin{equation}
g_{WZ}=-{3g^2\over 8\pi^2 f_\pi}.
\end{equation}
With an ideal mixing between $\omega$ and $\phi$ we have only $\phi-K^*-K$
interactions. The contribution to the phi self-energy is given by Fig.~1.c
and is written as
\begin{eqnarray}
\Pi^{(c)}_{\mu\nu}(k) &=& -g_{WZ}^2 T\sum\int{d^3l\over(2\pi)^3}
     {1\over (l^2-m_{K^*}^2)((l+k)^2-m_K^2)}
\nonumber\\[6pt]
&&\qquad\quad\times\epsilon_{\alpha\beta\gamma\mu}
                   \epsilon_{{\alpha'}{\beta'}{\gamma'}{\nu}}
                   l^\alpha k^\gamma
                   \left(g^{\beta{\beta'}}-l^\beta l^{\beta'}/m_{K^*}^2\right)
                   l^{\alpha'} k^{\gamma'}.
\end{eqnarray}

Finally, we include contributions from vector
meson loops. See Fig.~1.d and 1.e. 
These contributions have been neglected since 
there are Boltzmann suppression factors $\sim e^{-m_V/T}$
because of the large mass of vector mesons.
However, their contributions to phi mass are not negligible 
compared to the kaon effect. 
In the effective Lagrangian the phi meson couples to $K^*$ mesons by
the kinetic terms  
\begin{eqnarray}
{\cal L}_{VVV} &=& ig {\rm tr}
(\partial_\mu V_\nu - \partial_\nu V_\mu)[V^\mu, V^\nu],
\nonumber\\[6pt]
{\cal L}_{VVVV} &=& -g^2 {\rm tr} [V^\mu, V^\nu] [V^\mu, V^\nu].
\end{eqnarray}
Since the $K^*$ mass is almost twice that of kaon  
the contributions of $K^*$ mesons are suppressed by the Boltzmann factor, 
$N_{K^*}/N_K\sim e^{-m_K/T}\sim 0.1$ at $T= 200$ MeV. 
When we include the spin degeneracy factor
for vector mesons and a strong coupling constant with $K^*$ mesons, however,
we expect an effect with the same order of magnitude as that from kaons.
The importance of the $K^*$ meson has been pointed out in the calculation 
phi mass at finite temperature using QCD sum rules \cite{asakawa}.
The vector meson contribution to phi meson self-energy is given 
by 
\begin{eqnarray}
\Pi^{(d)}_{\mu\nu}(k) &=& -{1\over2}g^2 T\sum\int{d^3l\over(2\pi)^3}
     {1\over (l^2-m_{K^*}^2)((l+k)^2-m_{K^*}^2)}\cr
&&\qquad\quad\times\left[(2l_\mu+k_\mu)g_{\alpha\beta}
           -(l_\alpha-k_\alpha)g_{\beta\mu}
           -(l_\beta+2k_\beta)g_{\mu\alpha}\right]\cr
&&\qquad\quad\times\left[g^{\alpha{\alpha'}}
-(l^\alpha-k^\alpha)(l^{\alpha'}-k^{\alpha'})/m_{K^*}^2\right]
\left[g^{\beta{\beta'}}-l^\beta l^{\beta'}/m_{K^*}^2\right]\cr
&&\qquad\quad\times\left[(2l_\nu+k_\nu)g_{{\alpha'}{\beta'}}
           -(l_{\alpha'}-k_{\alpha'})g_{{\beta'}\nu}
           -(l_{\beta'}+2k_{\beta'})g_{\nu{\alpha'}}\right],
\nonumber\\[12pt]
\Pi^{(e)}_{\mu\nu}(k) &=& 2g^2T\sum\int{d^3l\over(2\pi)^3}
\left[3g_{\mu\nu}-{1\over m_{K^*}^2}(l^2g_{\mu\nu}-l_\mu l_\nu)\right]
{1\over(l^2-m_{K^*}^2)}. 
\end{eqnarray}

To obtain the effective mass of phi mesons we study the pole
position of the in-medium propagator in the limit where
the space momentum goes to zero, $\vec k\to 0$. 
In this limit we have $\gamma=0$, $\alpha=\beta$.  
The effective mass is determined by a solution of the equation 
\begin{equation}
k_0^2-m_\phi^2+\alpha(k_0,\vec k\to 0)=0.
\end{equation}
The temperature dependence of phi mass is shown in Fig. 2. 
We find that $\phi$
effective mass decreases with temperature 
and is reduced by about 20 MeV at $T=200$ MeV. 
The dominant contributions come from kaon loops
due to ${\cal L}_{\phi KK}$ and the $SU(3)_V$ symmetry breaking terms.
As temperature increases effects from vector mesons become important and
increase the effective mass of phi meson. 
As expected from our simple estimate their contributions are not negligible.
When we use phenomenological $\omega-\phi$ mixing we have 
$\rho-\pi$ loop contributions from Wess-Zumino terms which increase the
phi mass \cite{haglin}.

Compared with the calculation done by K. Haglin and C. Gale \cite{haglin}, 
we do not have the large effect from kaon tadpole loops which increases phi
mass. In our calculation we also have kaon
tadpole loop corrections (Fig.~1.b). However, these contributions
come from $SU(3)_V$ symmetry breaking terms and have different effect.
In the massive Yang-Mills approach which also 
includes vector mesons in chirally
symmetric way,  these tadpole loop contributions are exactly canceled by 
those from axial-vector-pseudoscalar loops \cite{song2}. 
This implies that chiral symmetry plays an important role in vector meson 
mass at finite temperature and shows different result.  
Calculations using QCD sum rules also show that the 
effective mass of phi mesons decreases with temperature.
However, we have a small effect compared to the result obtained from 
in Ref.~\cite{asakawa} and have an 
opposite result for $K^*$ contributions.
This difference requires more study in future. 

A small change in $\phi$ effective mass makes it hard to observe the 
double phi peak in dilepton spectrum 
which has been suggested as a possible probes of chiral phase transition 
in hot hadronic matter. Since it has 
been estimated that the effective width of phi
mesons in hot hadronic matter becomes about 
$20\sim 30$ MeV \cite{haglin2}, the mass shift obtained
from the calculation will be within the effective width.  
With very accurate mass resolution, we might have a chance to see
these small changes of the effective mass in dilepton spectrum. 
Details, however, will also depend on the coupling of phi meson 
to photon which will
be modified in hot hadronic matter as for $\rho$ mesons \cite{song3}. 
Further study in this direction is under way. 

Author is very grateful to Che Ming Ko for the discussion about 
the effect of $K^*$ mesons. 
He also thanks Volker Koch for useful conversations.
This work supported by the Director, Office of Energy Research, Office of High
Energy and Nuclear Physics, Division of Nuclear Physics, Division of Nuclear
Sciences, of 
the U. S. Department of Energy under Contract No. DE-AC03-76SF00098.

{\it Note added:} After finishing this work we learned of a recent work by
A. Bhattacharyya, et al. (nucl-th/9602042). For kaon loop effects they 
get similar result. However, the effects of vector mesons
and $\eta$ mesons are not included in their calculation.

\eject
\newpage

{\bf Figure captions}

{\bf Figure 1.:} One loop diagrams for the phi meson self-energy

{\bf Figure 2.:} Effective mass of phi meson at finite temperature


\begin{thebibliography}{40}

\bibitem{qgp} L. McLerran, Rev. Mod. Phys. 58 (1986) 1001.

\bibitem{lattice} J. B. Kogut and D. K. Sinclair,
                  Nucl. Phys. B 280 (1987) 625.

\bibitem{gl} T. Hatsuda and T. Kunihiro,
             Phys. Rev. Lett. 55 (1985) 158.\\
             J. Gasser and H. Leutwyler, Phys. Lett. B 184 (1987) 83.

\bibitem{pisaski} R. Pisarski, Phys. Lett. B 160 (1982) 222.

\bibitem{screen} C. DeTar, Phys. Rev. D 32 (1985) 276.\\
                 C. DeTar and J. Kogut, {\it ibid.} 36 (1987) 2828.

\bibitem{sum} A. I. Bochkarev and M. E. Shaposhnikov,
              Nucl. Phys. B268 (1986) 220.\\
              H. G. Dosch and S. Narison, Phys. Lett. B 203 (1988) 155.\\
              C. A. Dominguez and M. Loewe,
              Phys. Lett. B 233 (1989) 201.\\
              R. Furnstahl, T. Hatsuda and Su H. Lee,
              Phys. Rev. D 42 (1990) 1744.

\bibitem{dey} M. Dey, V. L. Eletsky and B. L. Ioffe,
              Phys. Lett. B 252 (1990) 620.

\bibitem{lee} T. Hatsuda, Y. Koike and Su H. Lee,
              Phys. Rev. D 47 (1993) 1225.

\bibitem{gale} C. Gale and J. I. Kapusta, Nucl. Phys. B 357 (1991) 65.

\bibitem{song1} Chungsik Song, Phys. Rev. D 48 (1993) 1375.

\bibitem{haglin} K. Haglin and C. Gale, Nucl. Phys. B 421 (1994) 613.

\bibitem{song2} Chungsik Song, Phys. Rev. D (1996) (to be published).

\bibitem{phi} Pin-Zhen Bi and Johann Rafelski, Phys. Lett. B 262 (1991) 485.

\bibitem{ko} M. Asakawa and C. M. Ko, Phys. Lett. B 322 (1994) 33.

\bibitem{hls} M. Bando, T. Kugo and K. Yamawaki, 
              Phys. Rep. 164 (1988) 217.

\bibitem{su3} A. Bramon, A. Grau and G. Pancheri, Phys. Lett. B 345 (1995) 263.

\bibitem{kapusta} K. Kajantie and J. Kapusta, 
                  Ann. Phys. (N.Y.) 160 (1985) 447.\\ 
                  U. Heinze, K. Kajantie, and T. Toimela, 
                  Ann. Phys. (N.Y.) 176 (1987) 218.


\bibitem{asakawa} M. Asakawa and C. M. Ko, Nucl. Phys. A 572 (1994) 732.

\bibitem{haglin2} C. M. Ko and D. Seibert, Phys. Rev. C 49 (1994) 2198.\\
                  K. Haglin, Nucl. Phys. A584 (1995) 719.
                 
\bibitem{song3} Chungsik Song, Su. H. Lee and C. M. Ko, 
                Phys. Rev. C 52 (1996) R476.

\end{thebibliography}
\end{document}